\documentclass[12pt,prd,preprint,superscriptaddress]{revtex4}
\begin{large}\end{large}\usepackage{revsymb}
\usepackage{amsmath}
\usepackage[dvips]{graphicx}
\newcommand{\be}{\begin{equation}}
\newcommand{\ee}{\end{equation}}
\newcommand{\ben}{\begin{eqnarray}}
\newcommand{\een}{\end{eqnarray}}

\begin{document}

\title{Non-adiabatic dark fluid cosmology}
\date{\today}

\author{W.S. Hipolito-Ricaldi\footnote{E-mail: whipolito@gmail.com}}
\affiliation{Universidade Federal do Esp\'{\i}rito Santo,
Departamento
de F\'{\i}sica\\
Av. Fernando Ferrari, 514, Campus de Goiabeiras, CEP 29075-910,
Vit\'oria, Esp\'{\i}rito Santo, Brazil}

\author{H.E.S.
Velten\footnote{E-mail: velten@cce.ufes.br}}
\affiliation{Universidade Federal do Esp\'{\i}rito Santo,
Departamento
de F\'{\i}sica\\
Av. Fernando Ferrari, 514, Campus de Goiabeiras, CEP 29075-910,
Vit\'oria, Esp\'{\i}rito Santo, Brazil}

\author{W. Zimdahl\footnote{E-mail: zimdahl@thp.uni-koeln.de}}
\affiliation{Universidade Federal do Esp\'{\i}rito Santo,
Departamento
de F\'{\i}sica\\
Av. Fernando Ferrari, 514, Campus de Goiabeiras, CEP 29075-910,
Vit\'oria, Esp\'{\i}rito Santo, Brazil}

\begin{abstract}
We model the dark sector of the cosmic substratum by a viscous fluid with an equation of state $p
= - \zeta \Theta$, where $\Theta$ is the fluid-expansion scalar and $\zeta$ is the coefficient of bulk viscosity for which we assume a dependence
$\zeta \propto \rho^{\nu} $ on the energy density $\rho$. The homogeneous and isotropic background dynamics coincides with that
of a generalized Chaplygin gas with equation of state $p = - A/\rho^{\alpha}$.
The perturbation dynamics of the viscous model, however, is intrinsically non-adiabatic and qualitatively different from the Chaplygin-gas case. In particular, it avoids short-scale instabilities and/or oscillations which apparently have ruled out unified models of the Chaplygin-gas type.
We calculate the matter power spectrum and demonstrate that the non-adiabatic model is compatible with the data from the 2dFGRS and the SDSS  surveys. A $\chi^{2}$-analysis shows, that for certain parameter combinations the viscous-dark-fluid (VDF) model is well competitive with the $\Lambda$CDM model.
These results indicate that \textit{non-adiabatic} unified models can be seen as potential contenders for a General-Relativity-based description of the cosmic substratum.
\end{abstract}

\maketitle

\section{Introduction}
\label{Introduction}

Most current cosmological models rely on the assumption that the dynamics of the Universe is described by Einstein's General Relativity (GR) and a material content that
is dominated by two so far unknown components, pressureless dark matter (DM) and dark energy (DE), a substance equipped with a large negative pressure. For reviews of the actual situation see \cite{rev,pad,dumarev} and references therein.
The preferred model is the $\Lambda$CDM model which also plays the role of a reference model for alternative approaches to the DE problem. While the $\Lambda$CDM model can describe most of the observations, there still remain puzzles \cite{peri}.
Other attempts to describe the apparently observed accelerated expansion of the Universe (see, however, \cite{sarkar}) are a potential
back-reaction mechanism from non-linear structure formation \cite{buchert} or models of the Lema\^{\i}tre-Tolman-Bondi type which are isotropic but not homogeneous \cite{enquist}.
Then there exists a line of investigation that modifies GR with the aim to obtain an accelerated expansion of the Universe as a result of the (modified) geometrical sector instead of matter with a negative pressure \cite{capofranca,duma}.
Here we focus on a class of approaches within GR that do not separate DM and DE from the start but regard the dark sector as a one-component substratum which exhibits properties of both DE and DM
on a joint footing. The best known models of this kind are Chaplygin-gas type cosmologies.
Starting with \cite{pasquier}, there has been a considerable activity in this field \cite{fabris1,bertolami,avelino1,bilic,Finelli1,zimdahl,gorini,NeoN,avelino}.
Among the host of models proposed over the years for the dark sector, the unified models
are minimal in the sense that they assume just one component that describes both dark matter and dark energy. Both components manifest themselves observationally only through their gravitational action. Therefore, a unified description is certainly attractive, at least as long there is no direct, i.e., other than gravitational, detection of either or even both of the dark components.
While the Chaplygin-gas models could well describe the SNIa results \cite{colistete}, i.e., the cosmic background dynamics, they seem to have fallen out of favor because of apparent problems to reproduce the matter power spectrum.
The difficulties of the generalized Chaplygin-gas cosmologies are related to the values of the sound speed of these models. Depending on the $\alpha$-parameter, the sound speed is either finite, i.e., it becomes  of the order of the speed of light, or its square is negative. In the first case, the small-scale perturbation behavior is oscillatory, in the second case there appear instabilities. In neither of these cases, the observed matter power spectrum is reproduced.
This circumstance has led the
authors of \cite{Sandvik} to the conclusion that Chaplygin-gas
models of the cosmic medium are ruled out  as competitive
candidates. Similar results were obtained from the analysis of the
anisotropy spectrum of the cosmic microwave background
\cite{Finelli2,Bean}, except possibly for low values of the Hubble
parameter \cite{bento3}. However, these conclusions rely on the
assumption of an adiabatic cosmic medium. It has been argued that
there might exist entropy perturbations, so far not taken into
account, which may change the result of the adiabatic perturbation
analysis \cite{NJP,ioav}. A problem here is the origin of
non-adiabatic perturbations which should reflect the internal
structure of the cosmic medium. The latter is unknown but it may
well be more complicated then suggested by the usually applied
simple (adiabatic) equations of state. Non-adiabatic
perturbations will modify the adiabatic sound speed. The speed of
sound has generally attracted interest as a tool to discriminate between
different dark energy models \cite{Erickson,DeDeo,Bean2,Weller,Raul}.
``Silent" Chaplygin gases were postulated by introducing ad hoc a non-adiabatic counter-term to exactly compensate the adiabatic pressure contribution \cite{ioav,amendola}.

Another option for a unified description of the dark sector are viscous models.
From early universe cosmology it is known that a bulk viscosity of the cosmic medium can induce an  inflationary phase \cite{murphy}.
A bulk viscous pressure in the early universe can be
the result of cosmological particle production \cite{zeldo,John}.
Under the conditions of
spatial homogeneity and isotropy, a scalar bulk viscous pressure
is the only admissible non-equilibrium phenomenon. The cosmological
relevance of bulk viscous media has subsequently been investigated in some
detail for an inflationary phase in the early universe (see
\cite{Roy,WZ,RM,Z2nd} and references therein).
With accumulating evidence for our present Universe to be in a stage of accelerated expansion,
an effective bulk viscous pressure was discussed as one of the potential sources for this phenomenon as well. It was
argued in \cite{antif,NJP}, that such a pressure can
play the role of an agent that drives the present
acceleration of the Universe. The option of a
viscosity-dominated late epoch of the Universe with accelerated
expansion was already mentioned in \cite{PadChi}, long before the direct observational evidence through the SN Ia data. For a
homogeneous and isotropic universe, the $\Lambda$CDM model and the
(generalized) Chaplygin-gas models can be reproduced as special
cases of this imperfect fluid description \cite{NJP}. The possibility of using cosmological
observations to probe and constrain imperfect dark-energy fluids was investigated in \cite{mota1} and \cite{mota2}.

It is obvious that the bulk viscosity contributes with a negative
term to the total pressure and hence a dissipative fluid seems to
be a potential dark energy candidate (For a recent preprint see \cite{barrow}). However, it is expedient to repeat a cautionary remark.
In traditional non-equilibrium thermodynamics
the viscous pressure represents a (small) correction to the
(positive) equilibrium pressure. This is true both for the Eckart \cite{eckart}
and for the Israel-Stewart theories \cite{israela,israelb}. Here we shall admit the
viscous pressure to be the dominating part of the pressure. This
is clearly beyond the established range of validity of
conventional non-equilibrium thermodynamics. Non-standard interactions are required to support such type of
approach \cite{antif,NJP}. Of course, this reflects the
circumstance that dark energy is anything but a ``standard" fluid. There are suggestions that viscosity might have its
origin in string landscape \cite{She}.
To successfully describe
the transition to a phase of accelerated expansion, preceded by a
phase of decelerated expansion in which structures can form, it is
necessary that the viscous pressure is negligible at high
redshifts but becomes dominant later on.

Extending previous work (\cite{rose,BVM}), this paper  provides a detailed study of perturbations in a viscous fluid model of the dark sector which in the homogeneous and isotropic background coincides with the dynamics of a generalized Chaplygin gas.
The differences in the perturbation dynamics of both approaches are traced back to an inherent non-adiabatic behavior of the viscous model.
In a sense, the viscous dark-fluid model can be seen as a non-adiabatic Chaplygin gas.
It does not suffer from the shortcomings of the so far considered adiabatic Chaplygin-gas models, i.e., it predicts neither
(unobserved) oscillations nor instabilities. Moreover, the non-adiabatic behavior is part of the model
and no ad hoc introduced counter terms to the adiabatic sound speed are required.
The resulting power spectrum fits both the 2dFGRS as well as the SDSS data and, for certain parameter combinations, the $\chi^{2}$-value of the viscous model is better than the corresponding $\Lambda$CDM-value.

As in \cite{rose,BVM}, we shall describe the bulk viscous pressure by Eckart's
expression \cite{eckart} $p = - \xi u^{i}_{;i}$ (Latin indices run from 0 to 3), where
the (non-negative) quantity $\xi$ is the (generally not constant)
bulk-viscosity coefficient and $u^{i}_{;i}$ is the fluid-expansion scalar which in the homogeneous and isotropic background
reduces to $3 H$, where $H = \frac{\dot a}{a}$ is the Hubble
parameter and $a$ is the scale factor of the Robertson-Walker
metric. By this assumption we ignore all the problems inherent in
Eckart's approach which have been discussed and resolved within
the Israel-Stewart theory \cite{israela,israelb} (see also
\cite{Roy,WZ,RM,Z2nd} and references therein). We expect that for
the applications we have in mind here, the differences are of
minor importance.

 The paper is organized as follows. Section \ref{Basic dynamics} introduces the basic dynamics of bulk viscous cosmology and describes the homogeneous and isotropic background solutions in analogy to
 their well-known (generalized) Chaplygin-gas counterparts. Section \ref{Perturbations} contains the general non-adiabatic perturbation dynamics. In order to point out the differences between the viscous and the Chaplygin-gas models, the dynamics is developed in parallel for both approaches.  In Section \ref{Numerical results} it is shown that the basic second-order equations only coincide deep in the matter-dominated era. Subsequently, the numerical solutions and the corresponding matter power spectra are presented and compared with the 2dFGRS and the SDSS data.
 A summary of our study is given in Section \ref{Summary}.

\section{Basic dynamics}
\label{Basic dynamics}
We assume the cosmic medium to be described by an energy-momentum
tensor
\begin{equation}
T^{ik} = \rho u^{i}u^{k} + p g^{ik}
 \,  \label{Tik}
\end{equation}
with the equation of state
\begin{equation}
p = - \zeta \Theta
 \,  \label{pTheta}
\end{equation}
of  a bulk-viscous
fluid,
where $\Theta \equiv u^{i}_{;i}$ is the expansion scalar and $\zeta$ is the coefficient of bulk viscosity.
In the homogeneous and isotropic background one has $\Theta = 3H$,
where $H$ is the Hubble rate. If, moreover, the background is
spatially flat, the Friedmann equation
\begin{equation}
3 H^{2} = 8\,\pi\,G\,\rho
 \,  \label{}
\end{equation}
implies $\Theta \propto \rho^{1/2}$. Let us further assume that $\zeta \propto \rho^{\nu}$.
This corresponds to a background equation of state
\begin{equation}
p = - A \rho^{\nu + 1/2}
 \,  \label{}
\end{equation}
with a constant $A > 0$. Comparing this with the equation of state of a generalized Chaplygin gas (subscript c),
\begin{equation}
p_{\mathrm{c}} = - \frac{A }{\rho^{\alpha}}
 \,  ,\label{}
\end{equation}
the correspondence $\alpha = -\left(\nu + \frac{1}{2}\right)$ is obvious. This will allow us to
apply known results from the Chaplygin-gas dynamics to the background dynamics of the viscous model.
The similarity between generalized Chaplygin gases and bulk-viscous fluids is well known \cite{Szydlowski,BVM}.
We recall that for $\nu = \frac{1}{2} \leftrightarrow \alpha = 1$ and for $A = 1$ both models contain the $\Lambda$CDM model as a special case. The traditional Chaplygin gas with $\alpha = 1$ is based in higher-dimensional theories \cite{jackiw}. The generalized Chaplygin gas with $0 < \alpha \leq 1$ has been related to a Born-Infeld type approach \cite{bertolami}.
The time dependence of the pressure is described by
\begin{equation}
\dot{p} = \left[\frac{\dot{\Theta}}{\Theta} + \nu \frac{\dot{\rho}}{\rho}\right]p
 \quad \Rightarrow\quad
\frac{\dot{p}}{\dot{\rho}} = \frac{p}{\rho}\left[\frac{\dot{\Theta}}{\Theta} \frac{\rho}{\dot{\rho}}
+ \nu\right]
 \,  .\label{}
\end{equation}
With
\begin{equation}
\dot{\rho} = - \Theta \left(\rho + p\right)\ , \qquad \dot{\Theta}
= - \frac{\gamma}{2} \Theta^{2}\ , \qquad \gamma = 1 + \frac{p}{\rho}\ , \label{backg}
\end{equation}
the adiabatic sound speed can be written as
\begin{equation}
\frac{\dot{p}}{\dot{\rho}} = \frac{p}{\rho}\left[\frac{1}{2} + \nu\right]
 \,  .\label{}
\end{equation}
For the energy density we have
\begin{equation}
\rho=\left[A+ B \left(\frac{a_{0}}{a}\right)^{\frac{3}{2}
\left(1- 2\nu\right)}\right]^{\frac{2}{1 - 2\nu}}
\quad \Rightarrow\quad H = \sqrt{\frac{8\,\pi\,G}{3}}\,\left[A+
B\left(\frac{a_{0}}{a}\right)^{\frac{3}{2}\left(1- 2\nu\right)}\right]^{\frac{1}{1 - 2 \nu}}\ .
 \label{rho}
\end{equation}
The deceleration parameter $q  = - 1 -
\frac{\dot{H}}{H^{2}}$ takes the form
\begin{equation}
q  =  -  \frac{1 - \frac{B}{2A}
\left(\frac{a_{0}}{a}\right)^{\frac{3}{2}\left(1- 2\nu\right)}}{1 +
\frac{B}{A}\left(\frac{a_{0}}{a}\right)^{\frac{3}{2}\left(1- 2\nu\right)}}\ .\label{qAB}
\end{equation}
Its present value $q_{0}$ is
\begin{equation}
q_{0} =   - \frac{1 - \frac{B}{2A}}{1 + \frac{B}{A}}\quad
\Leftrightarrow\quad \frac{B}{2A} = \frac{1 + q_{0}}{1 - 2q_{0}}\
.\label{q0}
\end{equation}
The value $a_{acc}$ at which the transition from decelerated to
accelerated expansion occurs, is given by
\begin{equation}
q = 0 \quad \Leftrightarrow\quad
\left(\frac{a_{acc}}{a_{0}}\right)^{\frac{3}{2}\left(1- 2\nu\right)} = \frac{B}{2A} \quad
\Leftrightarrow\quad \frac{a_{acc}}{a_{0}} =
\left(\frac{B}{2A}\right)^{\frac{2}{3\left(1- 2\nu\right)}} \ .\label{aq}
\end{equation}
Denoting the redshift parameter at which the
acceleration sets in by $z_{acc}$, provides us with
\begin{equation}
1 +  z_{acc} = \frac{a_{0}}{a_{acc}}  \quad \Rightarrow\quad
z_{acc} = \left(\frac{1-2q_{0}}{1 + q_{0}}\right)^{\frac{2}{3\left(1- 2\nu\right)}} - 1 \
.\label{zacc}
\end{equation}
In terms of $q_{0}$ the Hubble function in (\ref{rho}) becomes
\begin{equation}
\frac{H}{H_{0}} = \left(\frac{1}{3}\right)^{\frac{1}{1- 2\nu}}\left[1 - 2q_{0} + 2 \left(1 +
q_{0}\right)\left(\frac{a_{0}}{a}\right)^{\frac{3}{2}\left(1- 2\nu\right)}\right]^{\frac{1}{1- 2\nu}} \ ,
\label{H/H0q}
\end{equation}
while the corresponding energy density is ($3 H_{0}^{2} = 8 \pi G \rho_{0}$)
\begin{equation}
\frac{\rho}{\rho_{0}} = \left(\frac{1}{9}\right)^{\frac{1}{1- 2\nu}}\left[1 - 2q_{0} + 2 \left(1 +
q_{0}\right)\left(\frac{a_{0}}{a}\right)^{\frac{3}{2}\left(1- 2\nu\right)}\right]^{\frac{2}{1- 2\nu}} \ .
\label{r/r0q}
\end{equation}
For the equation of state parameter $\frac{p}{\rho}$ we
obtain
\begin{equation}
\frac{p}{\rho} = - \frac{1 - 2q_{0}}{1 - 2q_{0} + 2 \left(1 +
q_{0}\right)\left(\frac{a_{0}}{a}\right)^{\frac{3}{2}\left(1- 2\nu\right)}}
 \ ,
\label{p/rq}
\end{equation}
which implies
\begin{equation}
\gamma = 1 + \frac{p}{\rho} =  \frac{2 \left(1 +
q_{0}\right)\left(\frac{a_{0}}{a}\right)^{\frac{3}{2}\left(1- 2\nu\right)}}{1 - 2q_{0} + 2
\left(1 + q_{0}\right)\left(\frac{a_{0}}{a}\right)^{\frac{3}{2}\left(1- 2\nu\right)}}
 \ .
\label{gq}
\end{equation}
The same relations hold for the generalized Chaplygin gas with the replacement
$1 - 2\nu = 2(1+ \alpha)$.

\section{Perturbations}
\label{Perturbations}

For a pressure $p \propto -  \rho^{\nu} \Theta$ the corresponding linear perturbations, denoted by the hat symbol, are
\begin{equation}
\hat{p} = \left[\frac{\hat{\Theta}}{\Theta} + \nu \frac{\hat{\rho}}{\rho}\right]p
 \, .  \label{hp}
\end{equation}
Quantities without a hat are background quantities.
The perturbations (\ref{hp}) are non-adiabatic. Namely,
\begin{equation}
\hat{p} - \frac{\dot{p}}{\dot{\rho}} \hat{\rho} = p
\left(\frac{\hat{\Theta}}{\Theta} -
\frac{1}{2}\frac{\hat{\rho}}{\rho}\right) \neq 0
 \, .  \label{}
\end{equation}
Adiabatic perturbations are characterized by $\hat{p} =
\frac{\dot{p}}{\dot{\rho}} \hat{\rho}$. It is the difference from
$\hat{p} = \frac{\dot{p}}{\dot{\rho}} \hat{\rho}$ which makes the
perturbations non-adiabatic.
The expression (\ref{pTheta}) for the
pressure coincides with an equation of state $p = p(\rho) \propto -
\rho^{\nu + 1/2}$ only in the background. On the perturbative level,
Eq.~(\ref{pTheta}) cannot be reduced to $p = p(\rho)$.
Use of the relations (\ref{backg})
allows us to write
\begin{equation}
\frac{\hat{p}}{\rho + p} - \frac{\dot{p}}{\dot{\rho}}
\frac{\hat{\rho}}{\rho + p} = 3 H \frac{\dot{p}}{\dot{\rho}}
\left(\frac{\hat{\rho}}{\dot{\rho}} -
\frac{\hat{\Theta}}{\dot{\Theta}}\right)
 \, ,  \label{}
\end{equation}
or, with the abbreviations
\begin{equation}
P \equiv \frac{\hat{p}}{\rho + p} \ , \qquad D \equiv
\frac{\hat{\rho}}{\rho + p}\ ,\label{}
\end{equation}
\begin{equation}
P - \frac{\dot{p}}{\dot{\rho}} D = 3 H \frac{\dot{p}}{\dot{\rho}}
\left(\frac{\hat{\rho}}{\dot{\rho}} -
\frac{\hat{\Theta}}{\dot{\Theta}}\right)
 \, .  \label{P-}
\end{equation}
Both the combinations $P - \frac{\dot{p}}{\dot{\rho}} D$ on the
left-hand side and $\frac{\hat{\rho}}{\dot{\rho}} -
\frac{\hat{\Theta}}{\dot{\Theta}}$ on the right-hand side of
(\ref{P-}) are gauge-invariant, while the quantities $P$, $D$,
$\hat{\rho}$ and $\hat{\Theta}$ by themselves are not gauge-invariant.
Obviously, the basic quantities for the study of the non-adiabatic
perturbation dynamics are the energy-density perturbation
$\hat{\rho}$ and the perturbation $\hat{\Theta}$ of the expansion
scalar. This suggests starting with the first-order energy
conservation equation, while the perturbation of the expansion
scalar is governed by the first-order Raychaudhuri equation.

The general line element for scalar perturbations is
\begin{equation}
\mbox{d}s^{2} = - \left(1 + 2 \phi\right)\mbox{d}t^2 + 2 a^2
F_{,\alpha }\mbox{d}t\mbox{d}x^{\alpha} +
a^2\left[\left(1-2\psi\right)\delta _{\alpha \beta} + 2E_{,\alpha
\beta} \right] \mbox{d}x^\alpha\mbox{d}x^\beta \ .\label{ds}
\end{equation}
The perturbed 4-velocity is described by
\begin{equation}
\hat{u}^0 = \hat{u}_0  = - \phi \label{}
\end{equation}
and
\begin{equation}
a^2\hat{u}^\mu + a^2F_{,\mu} = \hat{u}_\mu \equiv v_{,\mu} \ ,
\label{}
\end{equation}
which defines the velocity perturbation $v$. A choice
$v=0$ corresponds to the comoving gauge. It
is also useful to introduce
\begin{equation}
\chi \equiv a^2\left(\dot{E} -F\right) \ .\label{}
\end{equation}
The combination $v + \chi$ is gauge-invariant.
It is convenient to describe the perturbation dynamics in terms of
gauge-invariant quantities which represent perturbations on
comoving (superscript $c$) hypersurfaces. These are defined as
\begin{equation}
\frac{\hat{\rho}^{c}}{\dot{\rho}} \equiv
\frac{\hat{\rho}}{\dot{\rho}} + v \ , \qquad
\frac{\hat{\Theta}^{c}}{\dot{\Theta}} \equiv
\frac{\hat{\Theta}}{\dot{\Theta}} + v \ , \qquad
\frac{\hat{p}^{c}}{\dot{p}} \equiv \frac{\hat{p}}{\dot{p}} + v \
 \, .  \label{defc}
\end{equation}
In our case we have
\begin{equation}
\frac{\hat{p}}{\dot{p}} = \frac{\frac{\hat{\Theta}}{\Theta} + \nu \frac{\hat{\rho}}{\rho}}{\frac{\dot{\Theta}}{\Theta} + \nu \frac{\dot{\rho}}{\rho}}
\qquad
\Rightarrow\qquad \frac{\hat{p}^{c}}{\dot{p}} = \frac{\frac{\hat{\Theta}^{c}}{\Theta} + \nu \frac{\hat{\rho}^{c}}{\rho}}{\frac{\dot{\Theta}}{\Theta} + \nu \frac{\dot{\rho}}{\rho}}
 .  \label{pcThetac}
\end{equation}
Recall that a constant bulk-viscosity coefficient corresponds to $\nu = 0$.
The perturbed energy balance may be written
\begin{equation}
\left(\frac{\hat{\rho}}{\rho + p}  -
3\psi\right)^{\displaystyle\cdot} + 3H \left(\frac{\hat{p}}{\rho +
p} - \frac{\dot{p}}{\dot{\rho}}\frac{\hat{\rho}}{\rho + p}\right)
+ \frac{1}{a^2}\left (\Delta v +\Delta \chi\right) =0 \ ,\label{ebpert}
\end{equation}
where $\Delta$ is the three-dimensional Laplacian.
From the momentum balance we have in first order
\begin{equation}
\frac{\hat{p}}{\rho + p} +\frac{\dot{p}}{\rho + p} v + \dot{v} +
\phi =0 \ .\label{mbpert}
\end{equation}
In terms of the quantities introduced in (\ref{defc}), the balances (\ref{ebpert}) and (\ref{mbpert}) may be combined into
\begin{equation}
\left(\frac{\hat{\rho}^c}{\rho + p} \right)^{\displaystyle\cdot} -
3 H\frac{\dot{p}}{\dot{\rho}}\frac{\hat{\rho}^c}{\rho + p} +
\hat{\Theta}^c =0 \ . \label{ebal}
\end{equation}
With
\begin{equation}
D^{c} \equiv \frac{\hat{\rho}^{c}}{\rho + p}\ ,\label{Dc}
\end{equation}
a more compact form of (\ref{ebal}) is
\begin{equation}
\dot{D}^{c} - 3H\,\frac{\dot{p}}{\dot{\rho}} \, D^{c}  +
\hat{\Theta}^c =0  \ . \label{dotD}
\end{equation}
The expansion scalar $\Theta $ is governed by the Raychaudhuri
equation
\begin{equation}
\dot{\Theta} + \frac{1}{3}\Theta^{2} + 2\left(\sigma^{2} -
\omega^{2}\right) - \dot{u}^{a}_{;a} - \Lambda + 4\pi\,
G\,\left(\rho + 3 p\right) = 0 \ . \label{}
\end{equation}
Up to first order, the perturbed Raychaudhuri equation can be
written in the form
\begin{equation}
\dot{\hat{\Theta}}^c + \frac{2}{3}\Theta\hat{\Theta}^c +
\frac{1}{a^{2}}\Delta P^{c} + \frac{\gamma}{6}\Theta^{2} \, D^{c}
= 0\ , \label{pertRay}
\end{equation}
where
\begin{equation}
P^{c} \equiv \frac{\hat{p}^{c}}{\rho + p}\ \label{defPc}\ .
\end{equation}
It is through the
Raychaudhuri equation that the pressure gradient comes into play.
The formulation of the perturbation dynamics in terms of
$\hat{\Theta}^c$ is particularly appropriate in the present case,
since via (cf. (\ref{pcThetac}) and (\ref{defPc}))
\begin{equation}
P^{c} = \frac{p}{\rho}\,\left[\frac{\hat{\Theta}^{c}}{\gamma\Theta} + \nu D^{c}\right]
\ ,\label{PcTc}
\end{equation}
the perturbation $\hat{\Theta}^c$ of the expansion scalar is
directly related to the pressure perturbation. Use of (\ref{PcTc})
in Eq.~(\ref{dotD}) provides us with a direct relation between the
pressure perturbations and the energy-density perturbations,
\\
\begin{equation}
P^{c} = - \frac{p}{\gamma \rho \Theta}\left[\dot{D}^{c} -
\Theta\,\left(\frac{p}{2\rho} + \nu \left(1 + 2 \frac{p}{\rho}\right)\right)\, D^{c}\right] \ .\label{PcDc}
\end{equation}
The pressure perturbation consists of a term which is proportional
to the energy-density perturbations $D^{c}$, but additionally of a
term proportional to the time derivative $\dot{D}^{c}$ of $D^{c}$.
Pressure perturbations are
not just proportional to the energy-density perturbations
as in the adiabatic case. There is an additional
dependence on the time derivative of the energy-density perturbations. The relation between pressure perturbations $P^{c}$ and
energy perturbations $D^{c}$ is no longer simply algebraic, equivalent to a
(given) sound-speed parameter as a factor relating the two. The
relation between them becomes part of the dynamics. In a sense,
$P^{c}$ is no longer a ``local" function of $D^{c}$ but it
is a function of the derivative $\dot{D}^{c}$ as well \cite{essay}. This is equivalent to
$\hat{p} = \hat{p}(\hat{\rho}, \dot{\hat{\rho}})$. It is only for
the background pressure that the familiar dependence $p = p(\rho)$
is retained.

Combining Eqs.~(\ref{dotD}), (\ref{pertRay}) and (\ref{PcDc}) and transforming to the $k$-space, we
obtain (using the same symbols as in the coordinate space) the second-order equation
\begin{eqnarray}
\ddot{D}^c &+& 3H\left[\frac{2}{3} - \frac{p}{2\rho}\left(1 + 2\nu\right) - \frac{1}{9}\frac{p}{\gamma
\rho}\,\frac{k^{2}}{H^{2} a^{2}}\right]\dot{D}^c  \nonumber\\ &-&
9H^{2}\left[\frac{1}{3}\left(\frac{\gamma}{2} + \frac{p}{\rho}\left(1 + 2\nu\right)\right) - \nu\gamma\frac{p}{2\rho}\left(1 + 2\nu\right)
- \frac{1}{9}
\frac{k^{2}}{H^{2} a^{2}}\frac{p}{\gamma\rho}\left(\frac{p}{2\rho} + \nu\left(1 + 2\frac{p}{\rho}\right)\right) \right]\, D^c = 0 \ .\nonumber\\
\label{ddDk}
\end{eqnarray}
It is obvious, that the pressure perturbations give rise to contributions both in the brackets that multiply $D^c$ and $\dot{D}^c$.
For comparison, we also write down the corresponding equation for the generalized Chaplygin gas (subscript c):
\begin{eqnarray}
\ddot{D}_{\mathrm{c}}^c &+& 3H\left[\frac{2}{3} +\alpha \frac{p}{\rho}\right]\dot{D}_{\mathrm{c}}^c  \nonumber\\ &-&
9H^{2}\left[\frac{\gamma}{6} - \alpha\left(1+\alpha\right) \gamma \frac{p}{\rho} - \frac{\alpha}{6}\frac{p}{\rho}\left(1- 3\frac{p}{\rho}\right) + \alpha \frac{p}{\rho}
\frac{k^{2}}{9 H^{2} a^{2}}
\right]\, D_{\mathrm{c}}^c = 0 \ .
\label{ddDkc}
\end{eqnarray}

\section{Numerical implementation and results}
\label{Numerical results}

For the numerical implementation it is convenient to use
\begin{equation}
\delta_{\mathrm{v}} \equiv \frac{\hat{\rho}^{c}}{\rho} = \gamma D^{c}\,
\ , \label{deltac}
\end{equation}
instead of $D^{c}$.
The subscript $\mathrm{v}$ stands for viscous and was introduced to distinguish the perturbations in our viscous dark-fluid model from those of the corresponding generalized Chaplygin gas.
In terms of the scale factor $a$, Eq.~(\ref{ddDk}) then takes the form
\begin{equation}
\delta_{\mathrm{v}}'' + f_{\mathrm{v}}\left(a\right)\delta_{\mathrm{v}}' + g_{\mathrm{v}}\left(a\right) \,\delta_{\mathrm{v}} = 0 \ ,\label{dddelkpr}
\end{equation}
where a prime denotes a derivative with respect to $a$ and the coefficients $f_{\mathrm{v}}(a)$ and $g_{\mathrm{v}}(a)$ are
\begin{equation}
f_{\mathrm{v}}\left(a\right) = \frac{1}{a}\,\left[\frac{3}{2} - 6\frac{p}{\rho} + 3\nu\frac{p}{\rho} - \frac{1}{3}\frac{p}{\gamma\rho}\,\frac{k^{2}}{H^{2}
a^{2}}\right]  \label{f}
\end{equation}
and
\begin{equation}
g_{\mathrm{v}}\left(a\right) = - \frac{1}{a^{2}}\,\left[\frac{3}{2}  + \frac{15}{2}
\frac{p}{\rho} - \frac{9}{2}\,\frac{p^{2}}{\rho^{2}}
- 9\nu\frac{p}{\rho} -
\left(\frac{1}{\gamma}\frac{p^{2}}{\rho^{2}} + \nu \frac{p}{\rho}\right)\frac{k^{2}}{H^{2}
a^{2}}\right]\ , \label{g}
\end{equation}
respectively. The quantities $H$, $\frac{p}{\rho}$ and $\gamma$ as
functions of $a$ are given in (\ref{H/H0q}), (\ref{p/rq}) and (\ref{gq}), respectively. The present value of the scale factor is set to $a_{0} = 1$ in the the numerical calculations.

The perturbation equation for the generalized Chaplygin gas that corresponds to (\ref{dddelkpr}) is
\begin{equation}
\delta''_{\mathrm{c}} + f_{\mathrm{c}} \left(a\right)\delta'_{\mathrm{c}}  + g_{\mathrm{c}} \left(a\right) \,\delta_{\mathrm{c}}  = 0 \ ,\label{dddelch}
\end{equation}
with
\begin{equation}
f_{\mathrm{c}}\left(a\right) = \frac{1}{a}\,\left[\frac{3}{2} - \frac{15}{2}\frac{p}{\rho} - 3\alpha\frac{p}{\rho}\right] \label{fch}
\end{equation}
and
\begin{equation}
g_{\mathrm{c}}\left(a\right) = - \frac{1}{a^{2}}\,\left[\frac{3}{2} + 12
\frac{p}{\rho} - \frac{9}{2}\,\frac{p^{2}}{\rho^{2}}
+ 9\alpha\frac{p}{\rho} + \alpha
\frac{p}{\rho}\,\frac{k^{2}}{H^{2}
a^{2}}\right]\ , \label{gch}
\end{equation}
respectively.
With $\alpha = - (\nu + \frac{1}{2})$ all the terms in (\ref{fch}) and (\ref{gch}) that are not multiplied by $k^{2}$ coincide with the corresponding terms in (\ref{f}) and (\ref{g}), respectively.
The quantities $H$, $\frac{p}{\rho}$ and $\gamma$  as functions of $a$ are also given by (\ref{H/H0q}), (\ref{p/rq}) and (\ref{gq}), respectively, with the replacement $1 - 2\nu = 2(1+ \alpha)$.
The quantities (\ref{f}), (\ref{g}), (\ref{fch}) and (\ref{gch}) are functions of the scale factor and depend on the parameters  $k$, $\nu$, $q_0$, $H_0$. We shall restrict ourselves to $\nu < \frac{1}{2}$.
We recall that sound propagation in the viscous fluid model is governed by a combination of the $k^{2}$ terms both in (\ref{f}) and in (\ref{g}). In the Chaplygin-gas model, on the other hand, the sound speed square is given by $-\alpha\frac{p}{\rho}$, the factor that multiplies the $k^{2}$ term in (\ref{gch}).
In contrast to the viscous fluid case there does not appear a $k^{2}$ term in (\ref{fch}).
Eq.~(\ref{dddelch}) with (\ref{fch}) and (\ref{gch}) reproduce the basic perturbation equations in
\cite{Finelli1} and \cite{Sandvik}.

At early times, i.e., for small scale factors $a \ll 1$, both the equations  (\ref{deltac}) and (\ref{dddelch}) coincide and take the asymptotic form
\begin{eqnarray}\label{EarlyViscous}
\qquad \qquad \delta'' + \frac{3}{2a} \,\delta' - \frac{3}{2a^2}\,\delta =0 \,,\qquad \qquad  (a \ll 1)
\end{eqnarray}
for all parameters $q_0$, $\nu$ and for all scales. Here, $\delta$ may either be $\delta_{\mathrm{v}}$ or $\delta_{\mathrm{c}}$.
The solutions of (\ref{EarlyViscous}) are
\begin{eqnarray} \label{asympsol}
\delta(a\ll1)= c_1 a + c_2 a^{-3/2} \,,
\end{eqnarray}
where $c_1$ and $c_2$ are integration constants. This means, at early times, the two models are indistinguishable. In particular, the non-adiabatic contributions to the viscous model
are subdominant on all scales. Moreover, for $a \ll 1$ we can also consider our model to be indistinguishable from the $\Lambda$CDM model.
This will allow us to follow the evolution of all models from the same initial conditions.
We shall use the fact that the matter power spectrum for the
$\Lambda$CDM model is well fitted by the BBKS transfer
function \cite{bbks}.
Integrating the $\Lambda$CDM model back from today to a distant past, say $z = 1.000$, we
obtain the shape of the transfer function at that moment.
The
spectrum determined in this way is then used as initial
condition for our viscous model. This procedure is similar to that described in
more detail in references \cite{sola,saulo}.
The numerical integration of Eq.~(\ref{dddelkpr}) and Eq.~(\ref{dddelch}) was
performed  with the help of the mentioned $\Lambda$CDM initial conditions where we
used the parameters of the WMAP5 and 2dFGRS best-fit data sets \cite{Lambda}.

\begin{figure}  
\hspace{0cm}
\includegraphics[width=16cm, height=18cm]{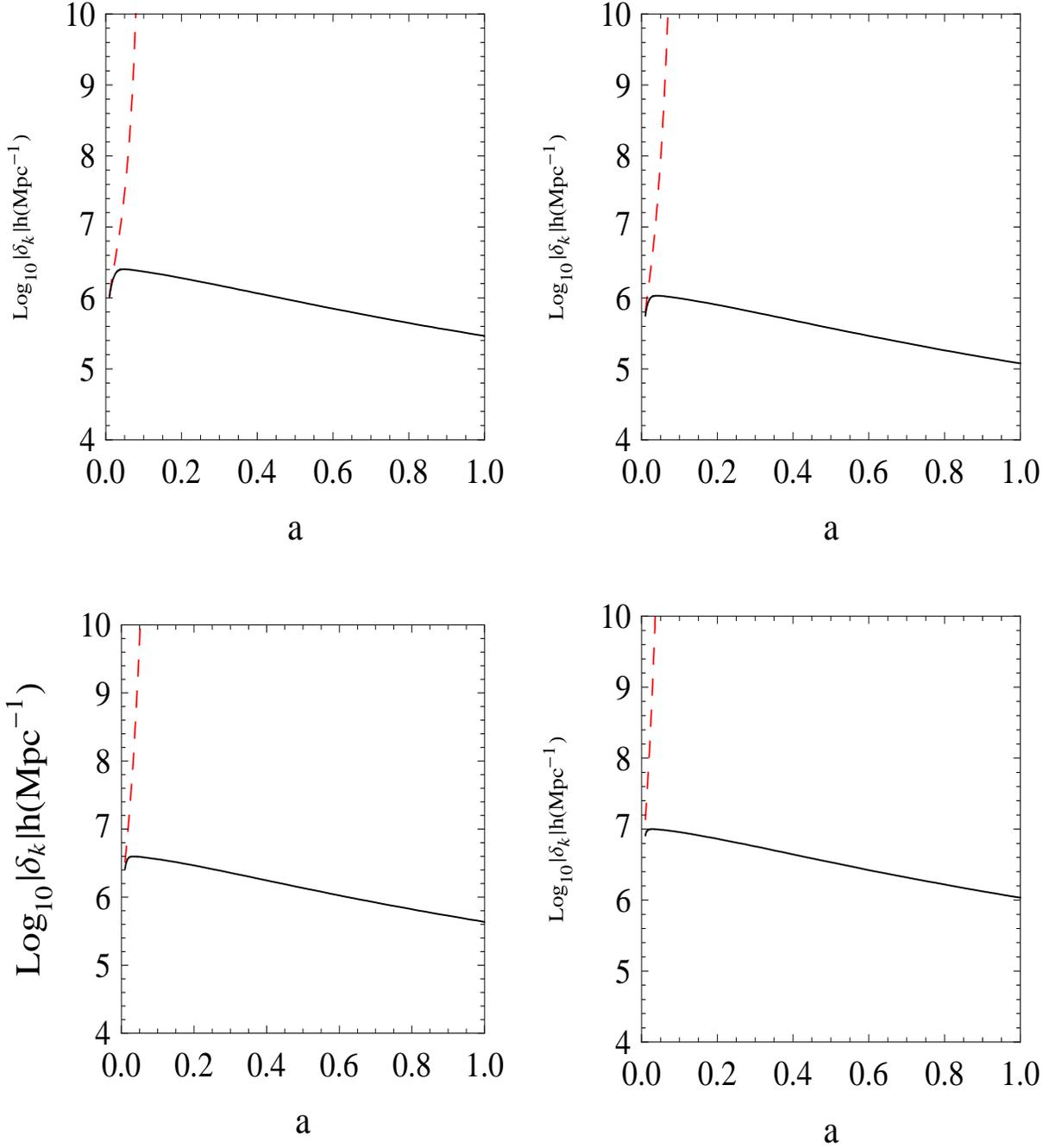}
\caption{\label{fig1}
Absolute values (logarithmic scale) of density fluctuations as function of the scale factor $a$
for $\nu=0$ ($\alpha=-1/2$) and $q_0=-0.5$ for different
scales. The values of $k$ are $k=0.5$ (top left), $k= 0.7$ (top right), $k =1$ (bottom left)
and $k =1.5$ (bottom right), all in units of $h Mpc^{-1}$. Solid curves represent the bulk viscous
model, dashed curves the corresponding GCG model.
Notice that both models are always different, except
at very early times.}
\end{figure}

\begin{figure}  
\hspace{0cm}
\includegraphics[width=16cm, height=18cm]{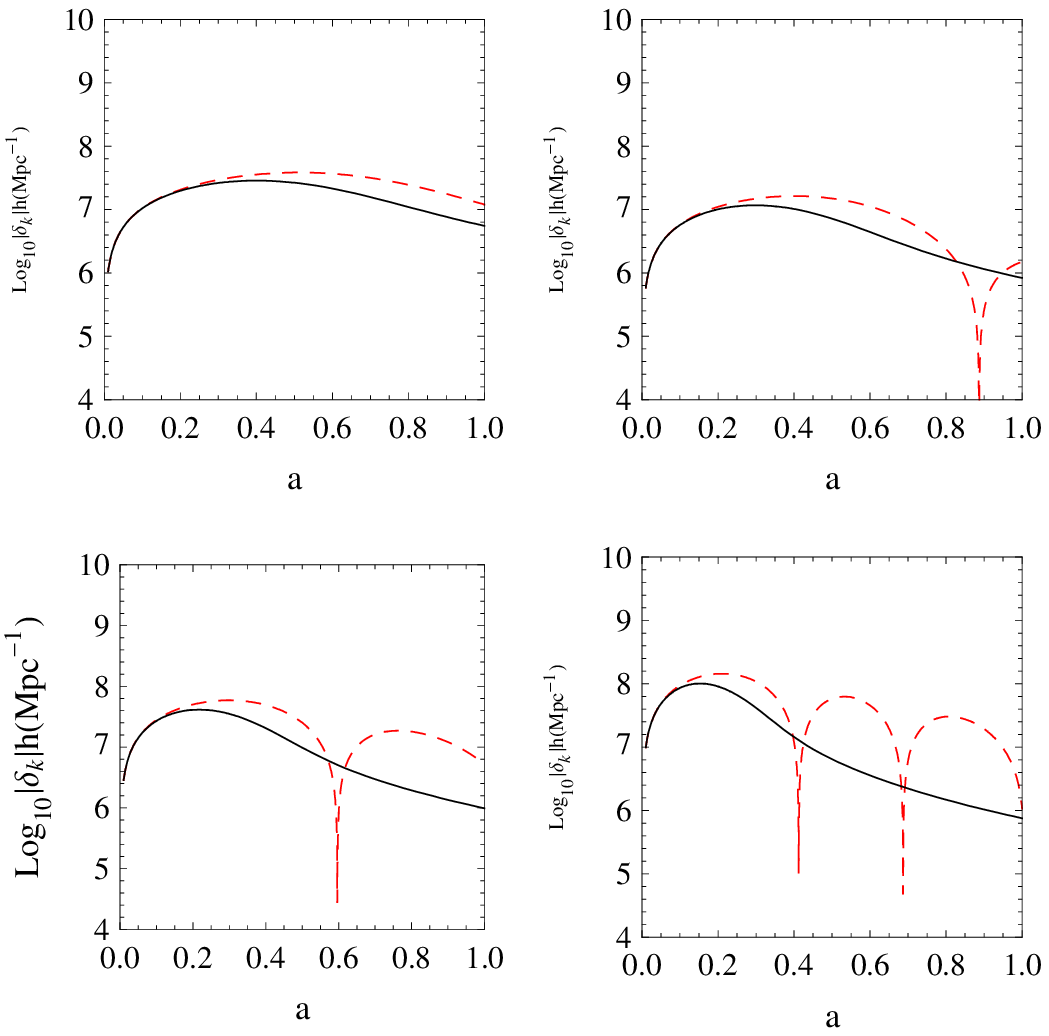}
\caption{\label{fig2}
Absolute values (logarithmic scale) of density fluctuations as function of the scale factor $a$
for $\nu=-1$ ($\alpha=1/2$) and $q_0=-0.5$  for different
scales. The values of $k$ are $k=0.5$ (top left), $k= 0.7$ (top right), $k =1$ (bottom left)
and $k =1.5$ (bottom right), all in units of $h Mpc^{-1}$. Solid curves represent the bulk viscous
model, dashed curves the corresponding GCG model.
Notice that both models are always different, except
at very early times.}
\end{figure}

In Figs.~\ref{fig1} and \ref{fig2} the density fluctuations for the viscous model are compared with those of the generalized Chaplygin-gas model for different values of the relevant parameters. Although identical in the background, both models are qualitatively very different at the perturbative level.
The density perturbations in the bulk-viscous scenario are  well behaved
at all times, while there appear instabilities or oscillations in the GCG model, depending on the parameter $\alpha$. The latter behavior was the main reason for discarding these models, except, possibly, for very small values of $\alpha$.
Fig.~\ref{fig1} shows the  blow up of
the Chaplygin-gas density for $\nu = 0$ ($\alpha =-1/2$). This reproduces a result of \cite{ioav}.
For $\nu = -1$ ($\alpha =1/2$), the Chaplygin-gas model predicts (unobserved) oscillations (cf. Fig.~\ref{fig2}), as was also found in \cite{ioav}. Neither of these unwanted properties hold
for our viscous model. This coincides with the results of \cite{BVM}.
Both models coincide for early times, confirming our previous analytical result, that non-adiabatic contributions are negligible in the past, but become relevant at a later period. The non-adiabatic contributions
are essential to avoid the mentioned unrealistic features of  Chaplygin-gas models.

\begin{figure} 
\hspace{0cm}
\includegraphics[width=14cm, height=22cm]{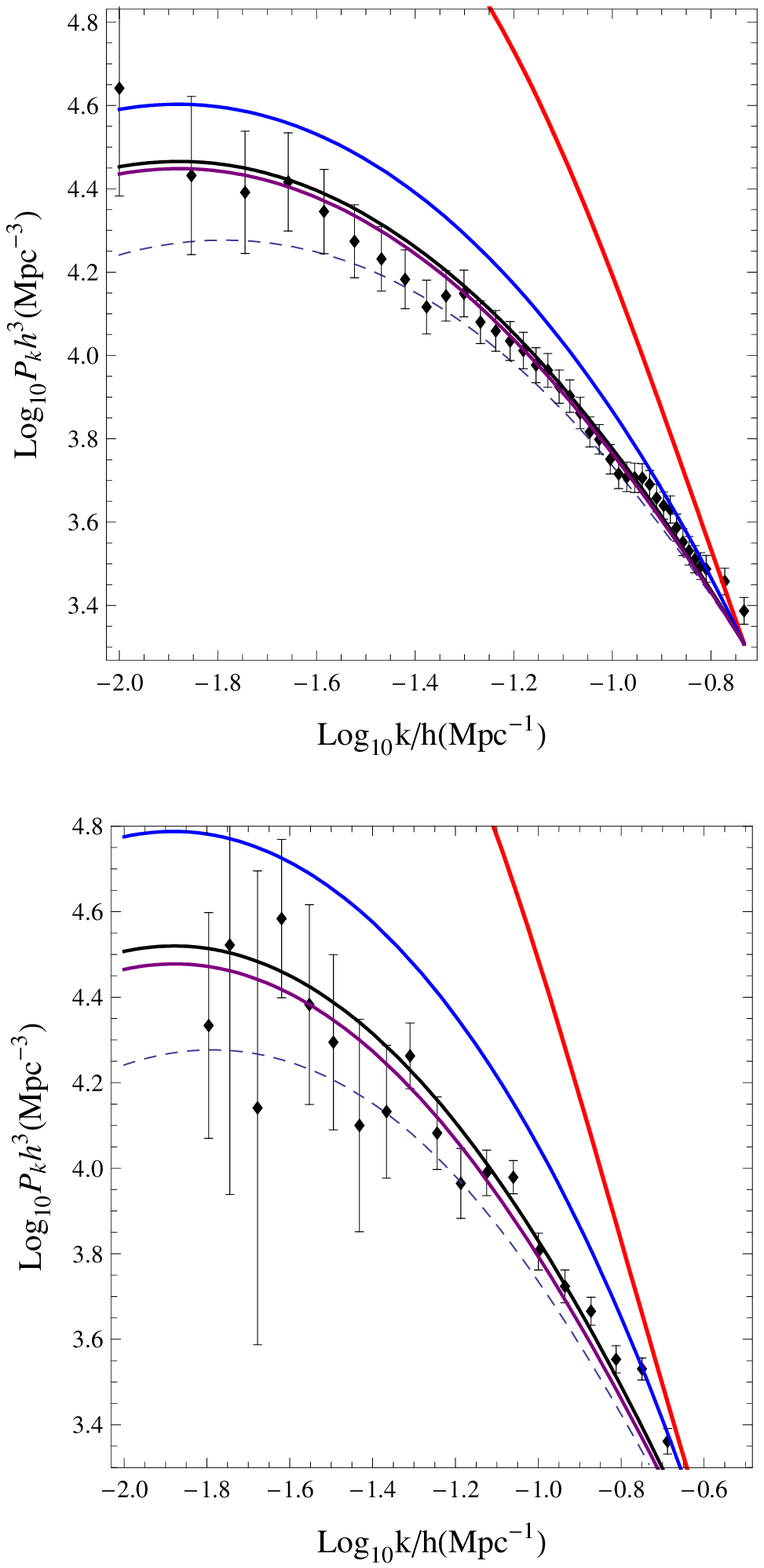}
\caption{\label{Spectrum025}
Density power spectrum for the bulk-viscous model with $\nu=0.25$ (solid curves)
and the $\Lambda$CDM model (dashed curves). From top to bottom the curves represent cases with $q_0=-0.4$,
$q_0=-0.2$, $q_0=0$ and $q_0=0.1$. The curves are compared with 2dFGRS data (top) and
and SDSS data (bottom).}
\end{figure}

\begin{figure} 
\hspace{0cm}
\includegraphics[width=14cm, height=22cm]{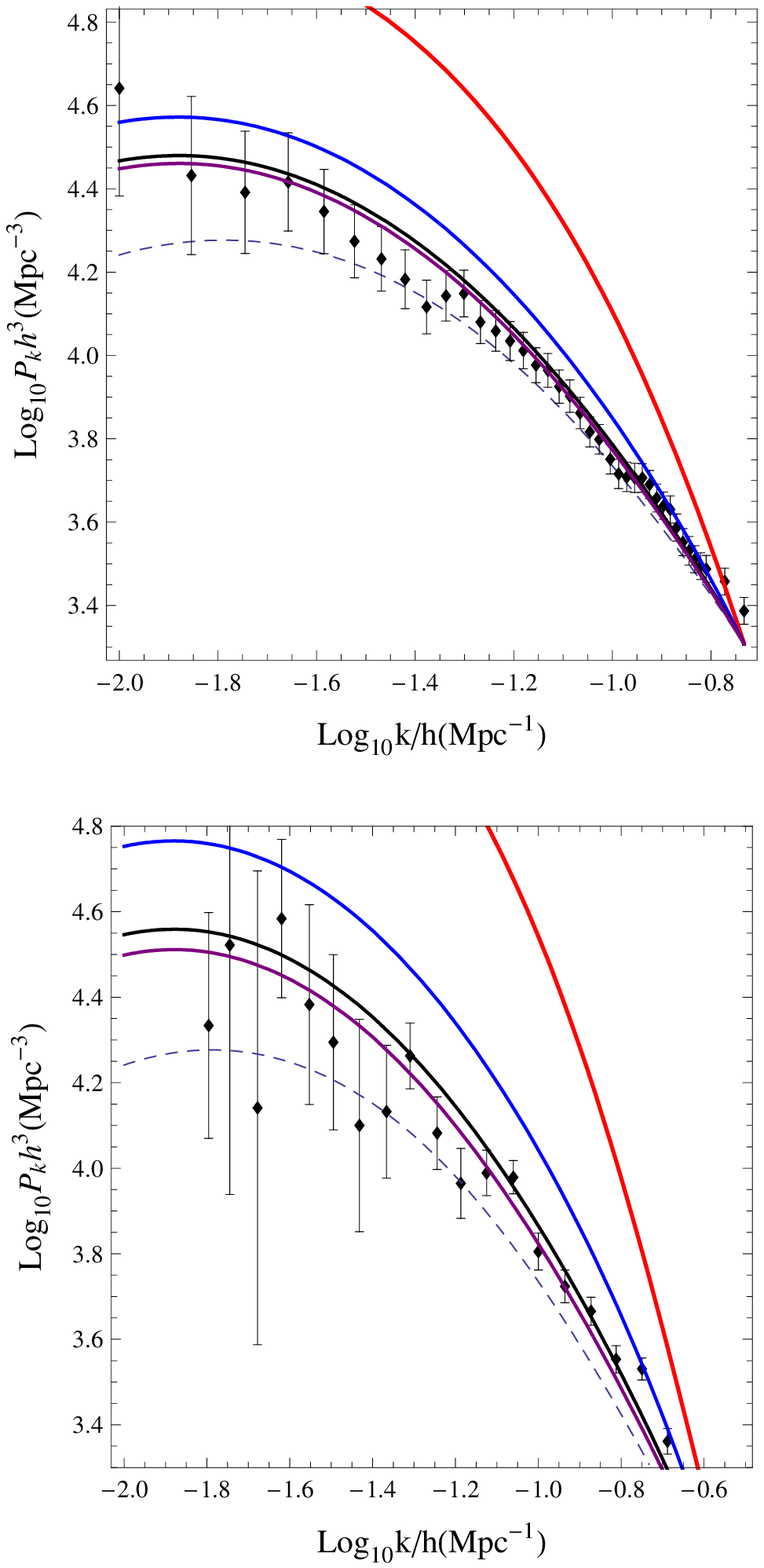}
\caption{\label{Spectrum0}
Density power spectrum for the bulk-viscous model with $\nu=0$ (solid curves)
and the $\Lambda$CDM model (dashed curves). From top to bottom the curves represent cases with $q_0=-0.4$,
$q_0=-0.2$, $q_0=0$ and $q_0=0.1$. The curves are compared with 2dFGRS data (top) and
and SDSS data (bottom).}
\end{figure}

\begin{figure} 
\hspace{0cm}
\includegraphics[width=14cm, height=22cm]{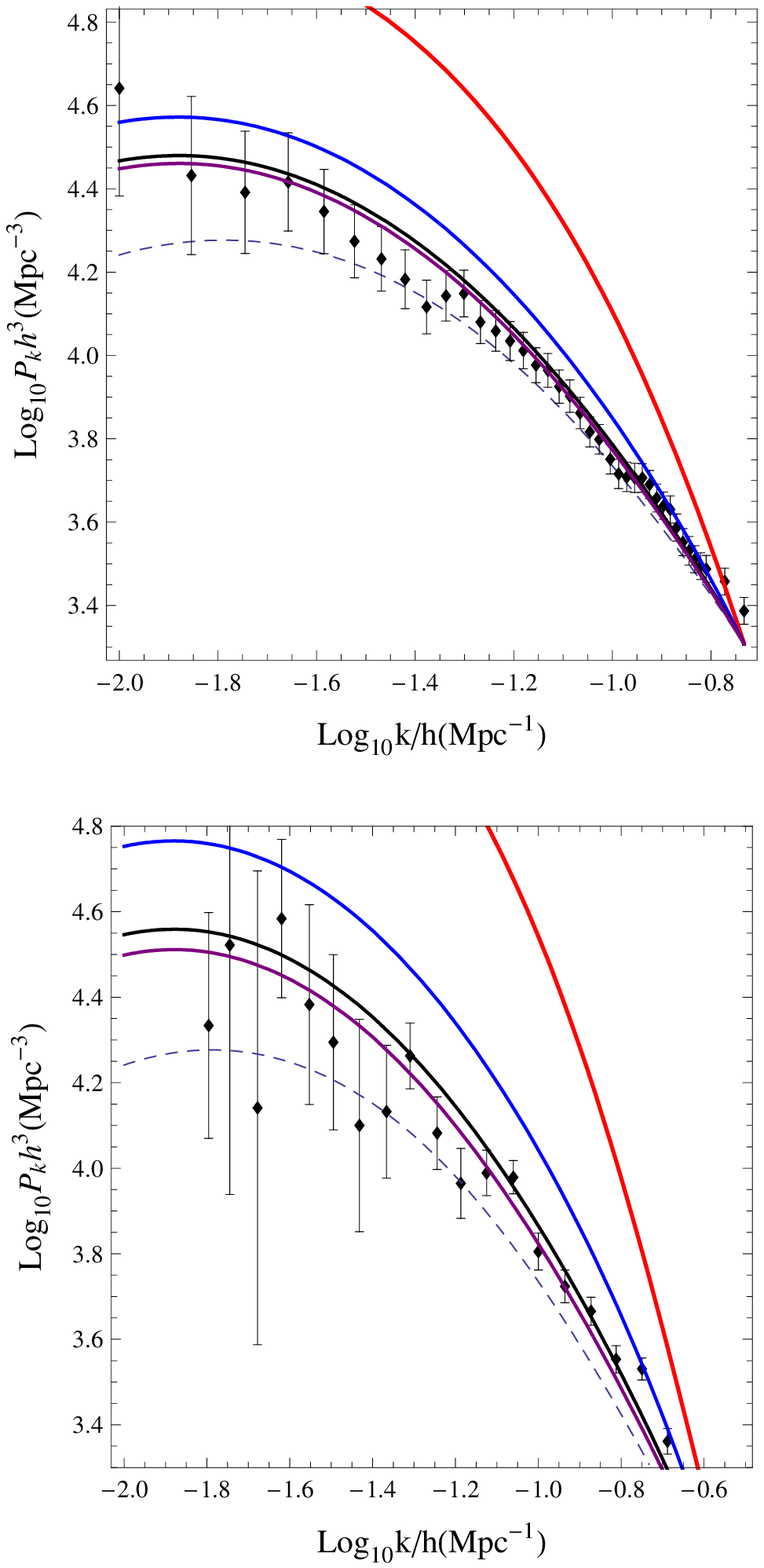}
\caption{\label{Spectrum-025}
Density power spectrum for the bulk-viscous model with $\nu=-0.25$ (solid curves)
and the $\Lambda$CDM model (dashed curves). From top to bottom the curves represent cases with $q_0=-0.4$,
$q_0=-0.2$, $q_0=0$ and $q_0=0.1$. The curves are compared with 2dFGRS data (top) and
and SDSS data (bottom).}
\end{figure}

\begin{figure} 
\hspace{0cm}
\includegraphics[width=14cm, height=22cm]{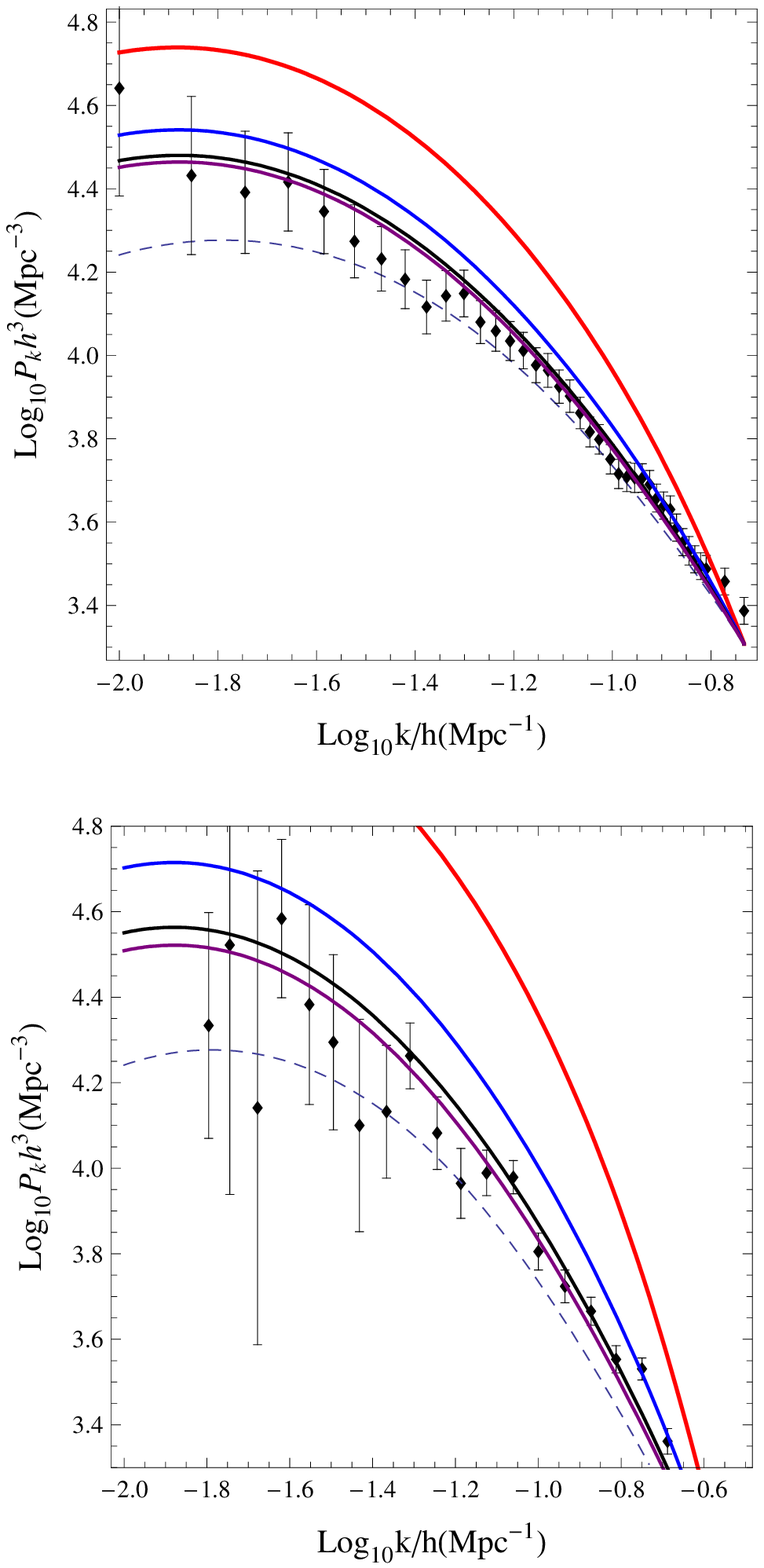}
\caption{\label{Spectrum-05}
Density power spectrum for the bulk-viscous model with $\nu=-0.5$ (solid curves)
and the $\Lambda$CDM model (dashed curves). From top to bottom the curves represent cases with $q_0=-0.4$,
$q_0=-0.2$, $q_0=0$ and $q_0=0.1$. The curves are compared with 2dFGRS data (top) and
and SDSS data (bottom).}
\end{figure}

\begin{figure} 
\hspace{0cm}
\includegraphics[width=14cm, height=22cm]{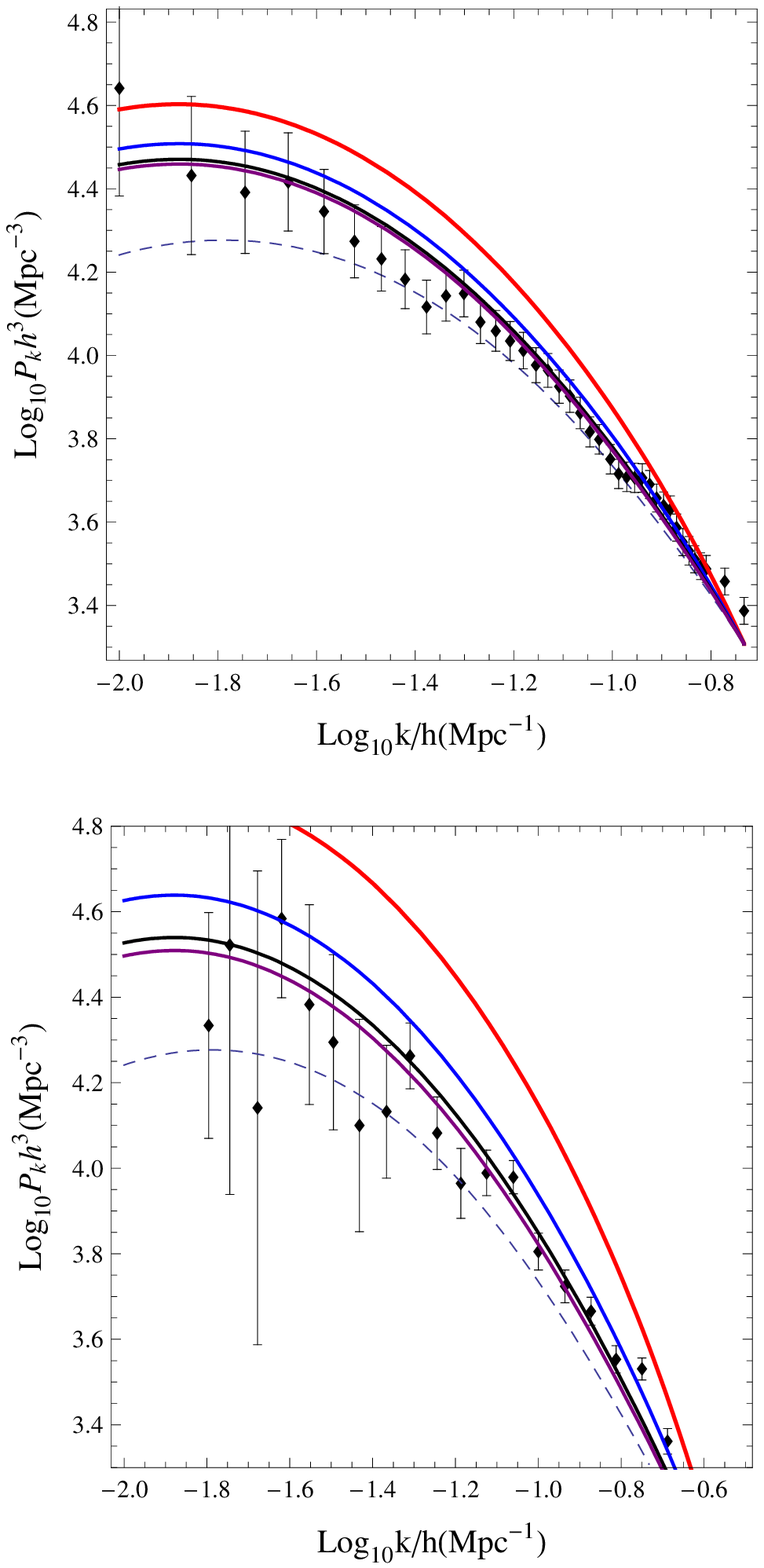}
\caption{\label{Spectrum-15}
Density power spectrum for the bulk-viscous model with $\nu=-1.5$ (solid curves)
and the $\Lambda$CDM model (dashed curves). From top to bottom the curves represent cases with $q_0=-0.4$,
$q_0=-0.2$, $q_0=0$ and $q_0=0.1$. The curves are compared with 2dFGRS data (top) and
and SDSS data (bottom).}
\end{figure}

\begin{figure} 
\hspace{0cm}
\includegraphics[width=14cm, height=22cm]{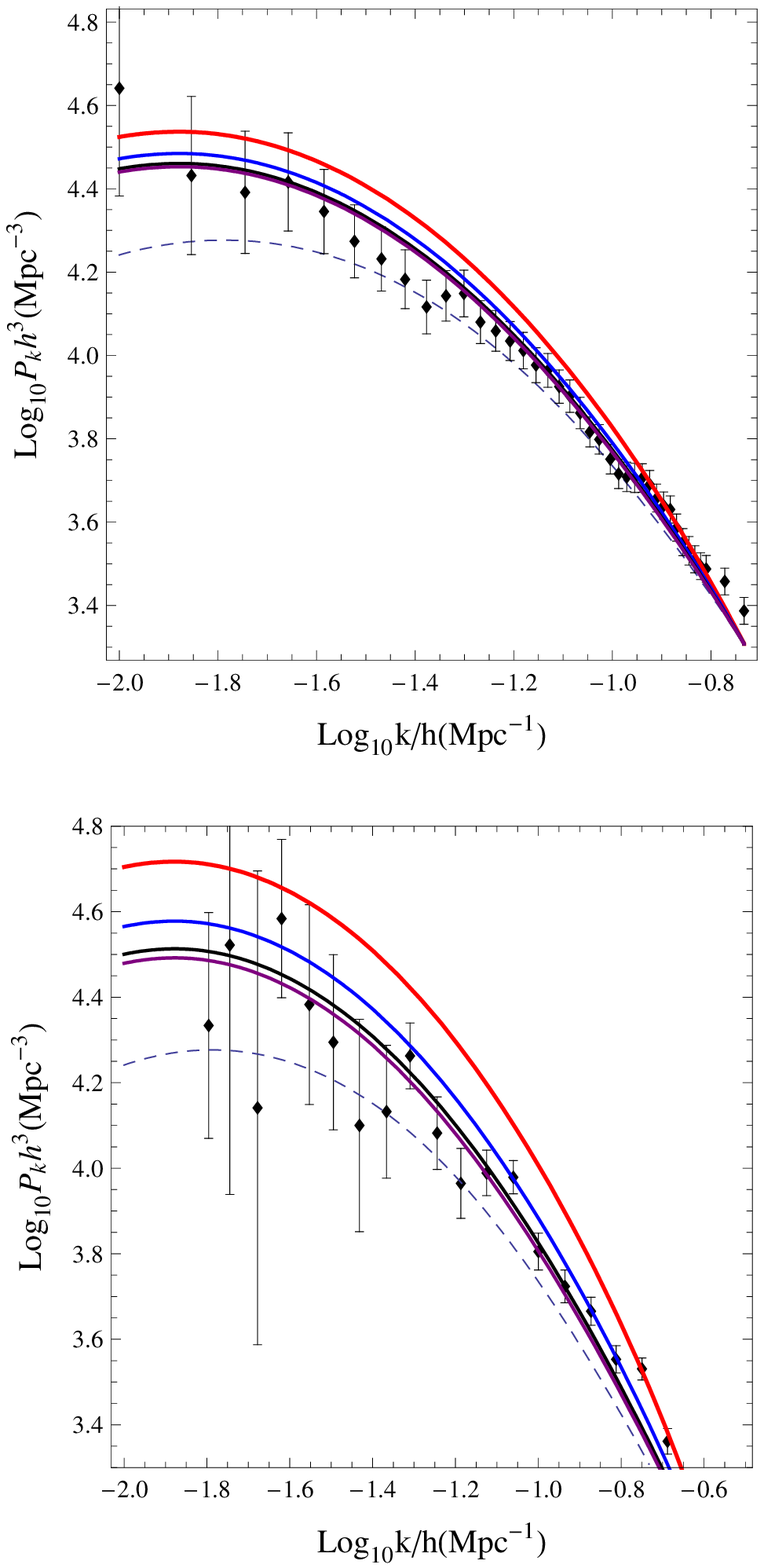}
\caption{\label{Spectrum-3}
Density power spectrum for the bulk-viscous model with $\nu=-3$ (solid curves)
and the $\Lambda$CDM model (dashed curves). From top to bottom the curves represent cases with $q_0=-0.4$,
$q_0=-0.2$, $q_0=0$ and $q_0=0.1$. The curves are compared with 2dFGRS data (top) and
and SDSS data (bottom).}
\end{figure}

\begin{figure} 
\hspace{0cm}
\includegraphics[width=14cm, height=22cm]{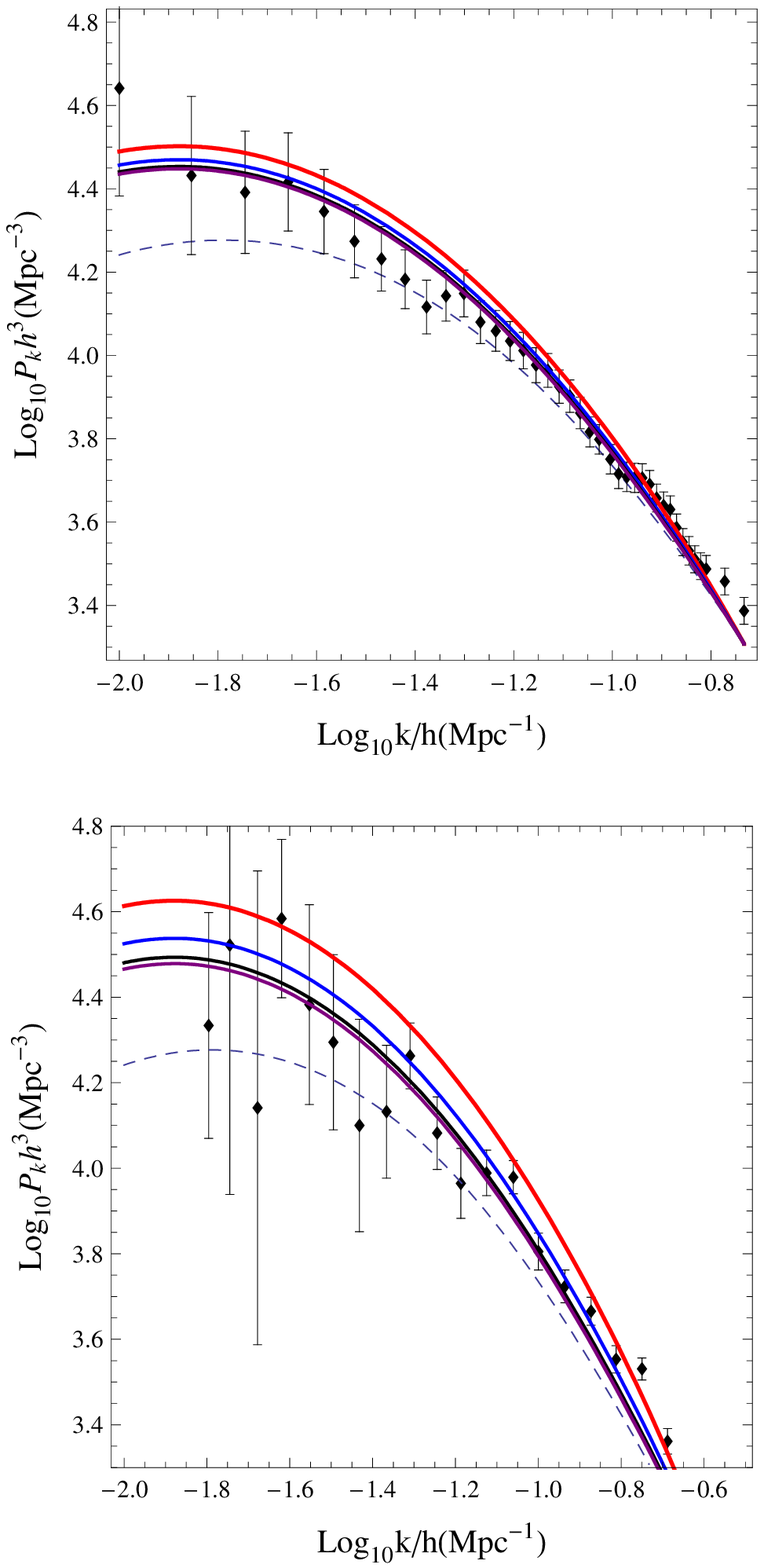}
\caption{\label{Spectrum-5}
Density power spectrum for the bulk-viscous model with $\nu=-5$ (solid curves)
and the $\Lambda$CDM model (dashed curves). From top to bottom the curves represent cases with $q_0=-0.4$,
$q_0=-0.2$, $q_0=0$ and $q_0=0.1$. The curves are compared with 2dFGRS data (top) and
and SDSS data (bottom).}
\end{figure}

The results for the matter power spectrum are shown and compared  with the 2dFRGS and SDSS samples for different parameters $\nu$ and $q_0$ in Figures~\ref{Spectrum025}-\ref{Spectrum-5}.
Two main features are observed here: (i) the bulk-viscous model is different from the $\Lambda$CDM
model for all parameter choices. In particular, this is also true for the case $\nu = - \frac{1}{2}$ (corresponding to $\alpha = 0$) which has a background equation of state $p \propto - \rho$.
(ii) For a certain range of the parameters $\nu$ and $q_0$, the model is in agreement with both the 2dFGRS  and SDSS data samples. There occur neither
oscillations nor instabilities. Negative values of $\nu$ are generally preferred. The more negative $\nu$ is, the more negative values of $q_{0}$ are compatible with the data. Large negative values of $\nu$ correspond to large positive values of $\alpha$. This is consistent with the results of recent studies on the perturbative behavior of generalized Chaplygin gases \cite{gorini,NeoN}, which prefer large values of $\alpha$ as well, although, on the other hand, pure Chaplygin-gas models suffer from causality problems for $\alpha > 1$.

Finally we perform a $\chi^2$-analysis and compare the values for the bulk-viscous model with the corresponding numbers for the $\Lambda$CDM model.
The results are summarized in Tab.~\ref{table1}. Our model turns out to
be competitive with the $\Lambda$CDM model for $q_0 \gtrsim -0.1$, with a minimum of $\chi^2$ around
$q_0 \sim 0$. It is expedient to point out, however, that the result of the comparison depends strongly on the priors and does not seem to be a good indicator for the quality of the model.



\begin{table*}
\caption{\label{table1}
Comparison of the $\chi^2$-values for the bulk-viscous model and the $\Lambda$CDM model
for different parameters $\nu$ and $q_0$.}\tiny{
\begin{tabular}{cccc} 
\hline 
\hline
$\nu$ &\hspace{0.5cm} $q_0$ & \hspace{0.5cm}$  \chi^2 \,(2dFGRS) $ & \hspace{0.5cm} $\chi^2 \,(SDSS)$
 \\
\hline
 \hline
0.25 & -0.3 & 2830.33 & 3776.76 \\
& -0.2  & 351.59&   459.76\\
&-0.1 &  75.07 &   72.89 \\
& 0 & 39.17&   54.79\\
  & 0.1  &  35.40 & 72.43  \\
&0.5& 37.12 & 98.76\\
 \hline
0 & -0.3 &  982.51& 2225.7 \\
& -0.2  & 238.79&  413.22 \\
&-0.1 & 85.97  &  100.08  \\
& 0 & 47.90&  51.42 \\
  & 0.1  & 37.40 &  56.36 \\
& 0.5& 37.12& 98.76\\
\hline
-0.25 & -0.3 & 570.22 &  1453.14\\
& -0.2  &183.90 &  331.18 \\
&-0.1 & 81.20  & 98.02   \\
& 0 & 48.92 & 52.39  \\
  & 0.1  & 38.44& 53.44  \\
& 0.5  & 37.12& 98.76 \\
\hline
-0.5 & -0.3 & 388.61 & 997.17 \\
& -0.2  &147.49 &  256.43 \\
&-0.1 & 75.76 & 87.27 \\
& 0 & 47.84& 51.74  \\
  & 0.1  & 38.57 &  53.08 \\
& 0.5  & 37.12 &  98.76 \\
\hline
-1.5 & -0.3 & 150.60 & 299.40 \\
& -0.2  &80.13 &  104.61 \\
&-0.1 & 52.88  &  56.75  \\
& 0 &41.54 & 50.06   \\
  & 0.1  & 36.96 &  56.69 \\
& 0.5  & 37.12 &  98.76 \\
\hline
-3 & -0.3 & 74.85 & 95.75 \\
& -0.2  &51.57 &  55.53 \\
&-0.1 & 41.64  &  49.91  \\
& 0 &37.34 & 55.27  \\
  & 0.1  & 35.69 &  64.08 \\
& 0.5  & 37.12 &  98.76 \\
\hline
$\Lambda$CDM & & 58.56 & 118.64 \\
\hline
 \hline
\end{tabular}}
\end{table*}

\section{Summary}
\label{Summary}

Unified models of the dark sector of the Universe may well be compatible with current observational data if this sector behaves as a bulk viscous fluid with a bulk-viscosity coefficient $\zeta \propto \rho^{\nu}$. In the homogeneous and isotropic background they coincide with generalized Chaplygin-gas models with $\alpha = -  \frac{1}{2} - \nu$, which are known to provide an adequate description of the SNIa data \cite{colistete}.
While Chaplygin-gas type cosmologies show a pathological behavior on the perturbative level, the corresponding bulk viscous model is well behaved and its matter power spectrum is compatible with the 2dFGRS and the SDSS observational data. It is the non-adiabatic character of the viscous fluid perturbation dynamics which is responsible for this difference. In a sense, our viscous model can be seen as a non-adiabatic generalized Chaplygin gas.
The model is observationally distinguishable from the $\Lambda$CDM model. Large negative values of $\nu$ are preferred ($\nu \lesssim -3$).
For certain parameter combinations a $\chi^{2}$-analysis favors our model over the $\Lambda$CDM model. However, we found a discrimination on this basis not sufficiently convincing since it depends strongly on the priors.
Our findings confirm and improve previous results on bulk-viscous cosmological models \cite{rose}.
But we consider the present study as preliminary, since it does not explicitly take into account a baryon component. A corresponding generalization is currently under investigation.
\vspace{1.0cm}


{\bf Acknowledgement}  We thank FAPES and CNPq (Brazil) for
financial support (Grants 093/2007 (CNPq and FAPES) and EDITAL
FAPES No. 001/2007).


%
\end{document}